Corresponding author:  Ranko Heindl
NIST ms 818.03
325 Broadway
Boulder, CO, 80305
Tel: 303-497-4224
Email: heindl@boulder.nist.gov

# Broadband Ferromagnetic Resonance Linewidth Measurement of Magnetic Tunnel Junction Multilayers

J. F. Sierra and F. G. Aliev

*Dpto. Fisica de la Materia Condensada, C-III, Universidad Autónoma de Madrid, 28049, Madrid, Spain*

R. Heindl, S. E. Russek, W. H. Rippard

*National Institute of Standards and Technology, Boulder, Colorado 80305*

*Work partially supported by US government, not subject to US copyright.

Abstract

The broadband ferromagnetic resonance (FMR) linewidth of the free layer of magnetic tunnel junctions is used as a simple diagnostic of the quality of the magnetic structure. The FMR linewidth increases near the field regions of free layer reversal and pinned layer reversal, and this increase correlates with an increase in magnetic hysteresis in unpatterned films, low frequency noise in patterned devices, and previous observations of magnetic domain ripple by use of Lorentz microscopy. Postannealing changes the free layer FMR linewidth indicating that considerable magnetic disorder, originating in the exchange-biased pinned layer, is transferred to the free layer.

Magnetic tunnel junctions (MTJs) are being developed for a variety of applications including magnetic recording sensors,[1] low-field magnetic sensors,[2] magnetic random-access memory (MRAM),[3] high-frequency spintronics devices,[4] and logic devices.[5] The quality of the MTJs depends sensitively on interfacial properties (see Refs. [6-7] for recent reviews). These properties include interfacial roughness, intermixing, and oxygen content, which in turn depend on materials used in the stack, deposition conditions, and annealing conditions. Ferromagnetic resonance (FMR) gives insight into both the magnetization dynamics and the quality of single ferromagnetic layers[8], multilayers[9], exchange-biased layers[10], and magnetic tunnel junctions[11]. The FMR linewidth depends on the intrinsic magnetic damping and extrinsic factors. The extrinsic factors include the quality of the ferromagnetic electrodes (e.g., the presence of local moments or antiferromagnetic phases produced by a non-ideal oxygen distribution near the tunnel barrier), leakage spin currents, thermally agitated spin diffusion, and the presence of undesirable magnetic coupling—such as Néel or "orange-peel" coupling. Both the intrinsic and extrinsic damping are important in determining switching times,[12] oscillator linewidths,[13] and detector bandwidths in MTJ devices.[4] The extrinsic damping may also be correlated with other important device properties, such as switching distributions in MRAM devices and low-frequency noise in sensors.

Here, we present data using broadband FMR to characterize interfacial properties of magnetic tunnel junctions at the wafer level and correlate regions of increased free-layer-linewidth with the reversal of the free and pinned layers, the appearance of magnetization ripple, and increased 1/f noise. In this paper we define linewidth as the FMR-linewidth of the free layer of the MTJ, and we have not measured the FMR response and the linewidth of the pinned layer of the MTJ. We show that there are considerable increases in the linewidth in field regions near the free-layer and pinned-layer reversal that correlate with peaks in the low frequency noise. The increase in the linewidth occurs in the same region where magnetization ripple is seen in Lorentz imaging studies. This increase in linewidth appears well before the reversal events, indicating that a disordered magnetic state begins before magnetostatic measurements indicate the layer reversal occurs. The correlation of the increase in linewidth with the presence of low-frequency magnetic noise indicates that they may have a common origin.

The tunnel junctions measured were grown in a high-vacuum sputtering chamber on insulating quartz wafers for the FMR studies and on oxidized silicon for the device measurements. The structure of the MTJ wafers used for FMR measurements was: [Ta(5nm)\Cu(5nm)\Ta(5nm)\Ru(2nm)\Ir$_{20}$Mn$_{80}$(10nm)\Co$_{90}$Fe$_{10}$(3nm)\Al$_2$O$_3$(1nm)\Co$_{60}$Fe$_{20}$B$_{20}$(2nm)\Ni$_{80}$Fe$_{20}$(23nm)\Ta(3nm)\Ru(7nm)] where Co$_{90}$Fe$_{10}$(3nm) forms the magnetic pinned layer. The wafers patterned into devices had a similar stack structure, except that the pinned layer was a Co$_{90}$Fe$_{10}$(2nm)\Ru(0.85nm)\Co$_{60}$Fe$_{20}$B$_{20}$(3nm) synthetic antiferromagnetic. The Co$_{60}$Fe$_{20}$B$_{20}$(2nm)\Ni$_{80}$Fe$_{20}$(23nm) forms the free-layer that is studied here. The free-layer thickness was optimized for low-field sensors and is thicker than typically used for MRAM or recording applications. All wafers were annealed at 250 °C in vacuum for 1 h in an applied magnetic field of 20 mT (these will be referred to "as deposited"). Some samples were further annealed in air for 1 h in an applied magnetic field of 100 mT along the same direction as the deposition field. In this study, we present data from four coupons obtained from the same MTJ FMR wafer: an as-deposited sample, a 250 °C annealed sample, a 275 °C annealed sample, and a 300 °C annealed sample, as well as data from one wafer with MTJ devices that are used for resistance noise measurements.

The high-frequency magnetization dynamics were measured by use of a commercial vector network analyzer (VNA) operating at frequencies up to 20 GHz. A VNA-FMR inductive technique was used to determine the FMR frequency and frequency-swept linewidth $\Delta f$ (full width at half maximum of the peak in the imaginary part of the susceptibility).[14] The FMR of the MTJ free layer was excited by a microwave magnetic field above a coplanar waveguide whose frequency was swept using the excitation signal from the VNA. The FMR spectra were obtained for different values of an applied in-plane magnetic field $H_{ap}$ (see Fig. 2(b) inset). The FMR spectra are normalized to a reference spectrum taken at a field such that the FMR resonance is outside the frequency range of interest. For high-frequency scans, the reference field was typically 0.6 T, whereas for low-field scans (< 10 GHz) the reference field was typically 0.25 T. The reference fields serve also as the initialization field, which resets the magnetization to a well defined state before each measurement. Both positive- and negative-going field sweeps were obtained in which the initialization/reference fields are anti-aligned and aligned with the pinned direction, respectively. Note that at each value of $H_{ap}$ a frequency swept FMR measurement was obtained, no field swept FMR spectra were measured. The absorption peaks were fit to a resonance model in which the real and imaginary parts were fit simultaneously.[14] The quasi-static magnetization characteristics were measured with a vibrating sample magnetometer (VSM). All measurements were performed at room temperature.

Figure 1 shows the resistance of a 64-element MTJ bridge as a function of applied magnetic field along with the resistance noise measured at 10 Hz (the applied voltage was $V_s$ = 1 V). The peaks in the resistance noise are due to thermally activated fluctuations of the magnetization at the tunnel barrier interface. Noise due to environmental magnetic field fluctuations was eliminated by use of a symmetric bridge, which is largely immune to environmental field noise, and by measuring in a shielded environment with low-noise field sources. Peaks in the low-frequency noise are seen in regions of the free layer reversal and the pinned layer reversal. The noise peaks are considerably larger when the sample starts with the pinned layer antiparallel to its pinned direction than when the sample starts with the pinned layer oriented parallel to its pinned direction. The defects giving rise to low-frequency magnetization noise are thought to be associated with the magnetization ripple induced in the free layer by disorder in the pinned layer.[15]

Figures 2a-c compare the magnetic field dependence of the quasi-static magnetization $M$, ferromagnetic resonance frequency $f_0$, and the FMR frequency-swept linewidth $\Delta f$ for the as-deposited and 275 °C annealed MTJ samples. As seen from the figures the annealing process changes a number of junction characteristics. The $M(H_{ap})$ curves show (Fig. 2a) that the annealing procedure sharpens the pinned-layer switching and pushes the switching to higher fields. Peaks in the free-layer linewidths are seen that are associated with the reversal of the pinned layer. After annealing, the increase in linewidth during reversal of the pinned layer is diminished, and the peak in the linewidth tracks the increase in switching field. A peak is also observed near the free layer reversal at low fields. There is a small reduction in this peak amplitude and a shift towards zero field after annealing.

Differences in the magnetization curves and in the curvature of the FMR frequency vs. magnetic field curves are seen after annealing (Fig. 2(a,b)), due to changes in saturation magnetization $M_s$ and anisotropy field $H_k$. Figure 3 plots the measured $M_s$ and $H_k$ values as functions of annealing temperature. These parameters were obtained by fitting the resonance frequency vs. magnetic field (Fig. 2b) to the Kittel equation $f_0 = (\gamma\mu_0/2\pi) [(H_k+M_s+H_{ap})\cdot(H_k+H_{ap})]^{1/2}$, where $\gamma = 1.76 \times 10^{11}$ s$^{-1}$T$^{-1}$ is the gyromagnetic ratio. Annealing decreases the anisotropy field and increases the saturation magnetization. The bulk of the free layer anisotropy arises from the coupling of the pinned layer and the free layer and reduction in anisotropy with annealing indicates a reduction in this coupling. The increase in magnetization may be due to the gettering of oxygen by the barrier and a subsequent increase in the magnetic moment of the adjacent layers. There is a significant change between the as-deposited sample, which undergoes a 250 °C *in situ* anneal, and the same sample after an additional *ex situ* 250 °C anneal. This may be due to the increased anneal time, a small error in the *in situ* temperature measurement, or the larger field used in the *ex situ* anneals.

The positive field region of the linewidth can be fit using a model that assumes the frequency swept linewidth is a sum of extrinsic broadening and intrinsic damping, multiplied by the change in resonant frequency with applied field[14]: $\Delta f = \left(\mu_0 H_0 + \dfrac{4\pi\alpha f_0}{\gamma}\right)\dfrac{df_0}{\mu_0 dH_{ap}}$, where $\alpha$ is the Gilbert damping parameter, $H_0$ is a measure of the variation in the local anisotropy field, and $\dfrac{df_0}{\mu_0 dH_{ap}} = \dfrac{\gamma}{2\pi}\sqrt{1+\left(\dfrac{\gamma\mu_0 M_s}{4\pi f_0}\right)^2}$. It is more convenient to fit a calculated effective field-swept linewidth $\Delta H_{eff} \equiv \Delta f / \dfrac{df_0}{dH_{ap}} = \mu_0 H_0 + \dfrac{4\pi\alpha f_0}{\gamma}$, since this should be a linear function of the resonant frequency and should normalize out the increase in $\Delta f$ observed near zero frequency due to the large change in resonance frequency with small changes in the local anisotropy field (dispersion effects).

We note that $\Delta H_{eff}$ is not identical to a measured field-swept linewidth and is a better method for characterizing MTJ devices. In a true field-swept linewidth measurement, the magnetic structure of the device changes during the field sweep and makes the resulting linewidth difficult to interpret. Fits to effective field swept linewidth as a function of resonant frequency are shown in the inset of Fig. 2c. The slope and intercept yield the values of $\alpha$ and $H_0$, respectively. $\Delta H_{eff}$ is not linear with resonant frequency, but instead increases significantly at low frequencies. This indicates that the increases in the linewidth do not solely result from static disorder and dispersion effects and a more realistic model of the disorder is needed to describe the data.[16] The values of $\alpha$ and $H_0$ obtained from fitting the positive low field region from 1 mT to 100 mT ($f_0$ up to 10 GHz) yield $\alpha$ = 0.0055 ± 0.0005 and $H_0$= 2.2mT ± 0.5 mT for the MTJ films. The sensitivity of $\alpha$ and $H_0$ to the selected fitting range makes quantitative comparison of these parameters for the different annealing procedures difficult. However, the linear fits show a consistent decrease in the disorder parameter $H_0$ of approximately 0.3 mT after the additional annealing.

A detailed comparison between the magnetic field-dependent magnetization and the linewidth reveals a strong correlation between these quasi-static and dynamic characteristics. In Figure 4 (a) we show the behavior of the as-deposited and 300 °C annealed MTJs near the region of the pinned layer reversal for the positive going field sweep. The data show an increase in the sharpness of the pinned layer reversal and a decrease in the amplitude of the linewidth peak upon ex-situ annealing., which results from a decrease in disorder within the pinned layer. During its reversal, the pinned layer provides a strong inhomogeneous magnetostatic field acting on the free layer, which in-turn gives rise to a large extrinsic linewidth. Decreasing the disorder in the pinned layer decreases these extrinsic linewidth broadening mechanisms and results in the observed decreased linewidth, as has been seen in other disordered systems.[17] Figure 4b shows the correlation between the linewidth and the quasi-static magnetization for the 275 °C annealed sample in more detail. Here both the positive- and negative-going branches of the field sweep are shown. The increase in the linewidth is larger during the positive-going sweep, which corresponds to starting with the pinned layer in its unfavorable orientation. This is similar to the low-frequency noise spectra and indicates that when the pinned layer starts out oriented in a direction opposite to its pinned direction, more disorder occurs during the reversal process.

The peak in the linewidth near the free layer reversal is fundamentally different from the peak during the reversal of the pinned layer. The peak near the free layer reversal is largely due to dispersion effects. In this region, a small amount of disorder gives rise to a large frequency-swept linewidth. The failure of the simple model (Fig. 2C inset), which assumes that the linewidth results from independent magnetic regions with static disorder, indicates that a more complicated analysis is needed to properly describe the linewidth broadening in MTJ devices. This correlates with Lorentz microscopy studies that show the magnetic structure, in similar MTJs, changes dramatically with the appearance of magnetization ripple in the field region before reversal occurs.[15] On the other hand, in the field region of the pinned layer reversal, the peak in the linewidth is mainly due to magnetostatic field inhomogeneities coming from the pinned layer.

In summary, we found a correlation between the regions of increased FMR free-layer linewidth and (1) the reversal regions of the free and pinned layer determined by quasi-static magnetization measurements, (2) the regions of excess 1/f noise, and (3) the regions of large magnetization ripple. The regions of increased linewidth, as well as the regions of increased 1/f noise, are history-dependent, i.e. depend on the starting state of the MTJ. Upon annealing, the peak in linewidth near the reversal of the pinned layer dramatically decreases and sharpens, while in the region of the free layer reversal, there is a modest decrease in disorder after annealing. The increase in FMR linewidth is a simple method for monitoring the disordered magnetic structure and the improvement created by the annealing process. Hence, the study of FMR linewidth of MTJs may be a useful diagnostic of the quality of the MTJs being developed for a variety of applications.

**Acknowledgments**

The authors thank Tony Kos for help with experiments. Support from the Spanish MEC (MAT2006-07196) is gratefully acknowledged. One of authors (JFS) also acknowledges support received from a Spanish FPI grant (BES-2004-5594) and the NIST Electromagnetics Division during his stay at Boulder during the summers of 2006 and 2007.

## Figure Labels

**Figure 1.** Resistance ($V_s$ divided by the total current passing through the bridge) of a symmetric MTJ bridge that consists of 64 (16 MTJs per bridge leg) 10 x 20 µm MTJs versus applied magnetic field: (a) positive- and (b) negative-going branch. Data on the resistance noise of the MTJ bridge for both positive- and negative-going branches are also plotted at 10 Hz. The insets show (a) the noise spectra $V_n$ obtained by taking the average of the Fourier transform magnitude of $V_{out}(t)$ at zero applied field and (b) the measurement geometry - see Ref. 18.

**Figure 2**. Magnetic field dependence of (a) magnetization (b) FMR frequency and (c) linewidth for the as-deposited and 275 °C annealed MTJs. The inset to (b) shows a frequency swept FMR spectra (imaginary part of the normalized transmitted signal) for the 275 °C annealed MTJ at $\mu_0 H_{ap}$ = 148 mT. The inset to (c) shows fits to the effective field-swept linewidth data for the as-deposited MTJ.

**Figure 3.** Anisotropy field $H_k$ and saturation magnetization $M_s$ as functions of annealing temperature determined from the fits of the FMR frequency versus magnetic field data. Data for the as-deposited MTJs, which are heated to 250 °C in the vacuum chamber, are also plotted.

**Figure 4.** (a) Comparison of magnetization loops and free-layer linewidths in the region corresponding to the switching of the pinned layer for as-deposited and 300 °C annealed MTJs. The solid arrow indicates the direction of the magnetic field sweep for all of the curves. (b) Correlation between the quasi-static magnetization loop and the linewidth for the 275 °C annealed sample. Both positive-going and negative-going curves are shown. The horizontal thin and thick arrows represent the directions of the free and pinned layer, respectively.

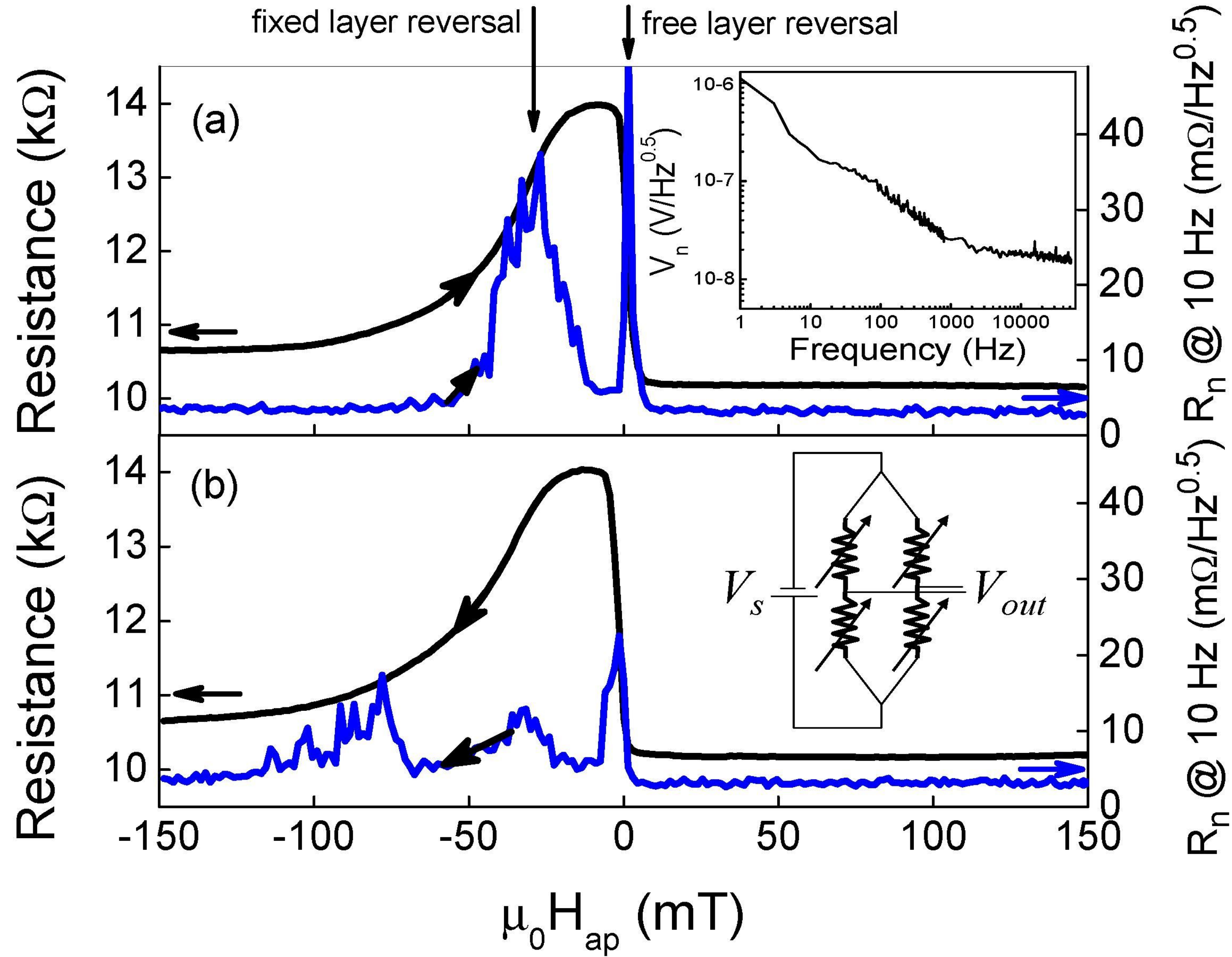

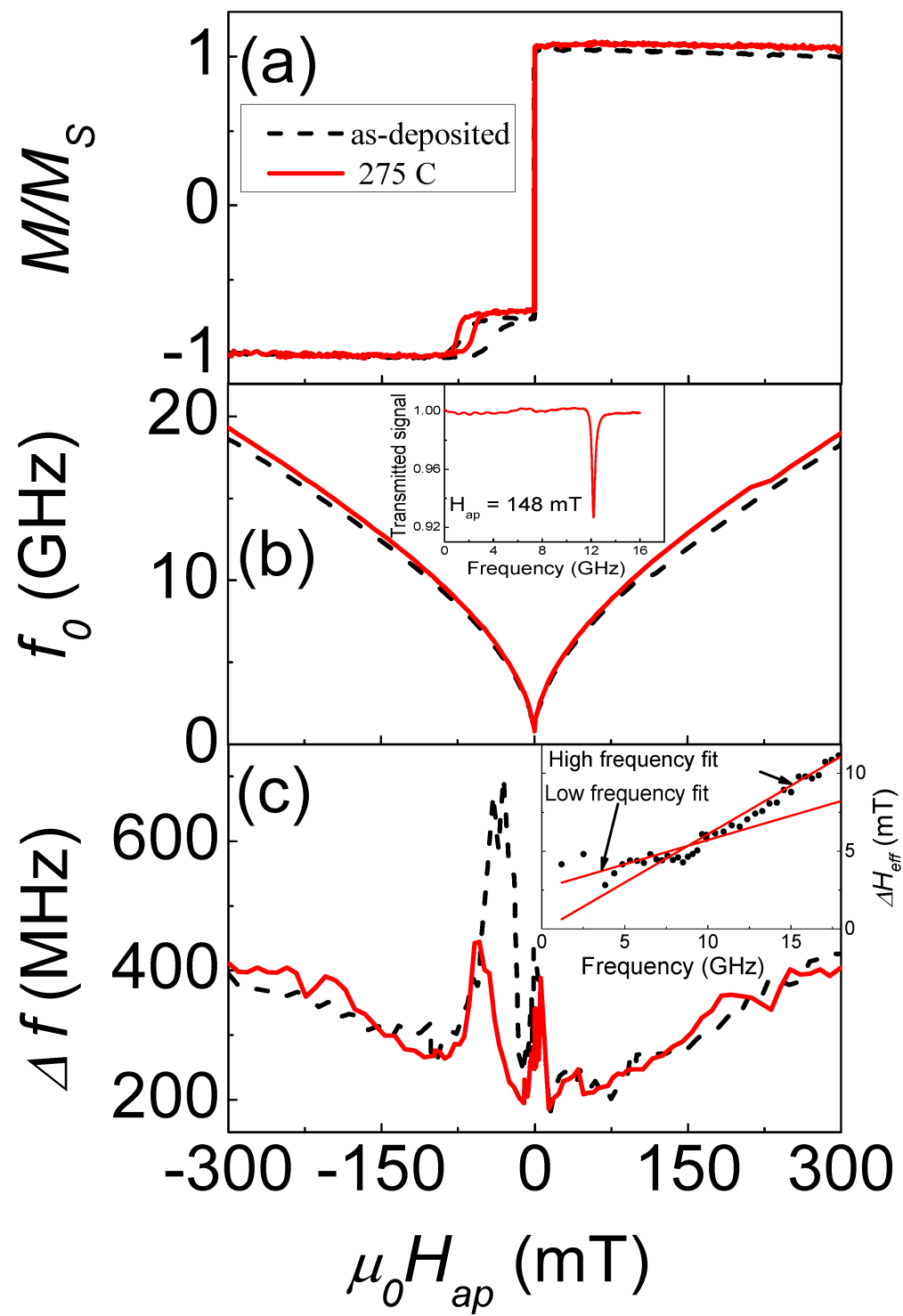

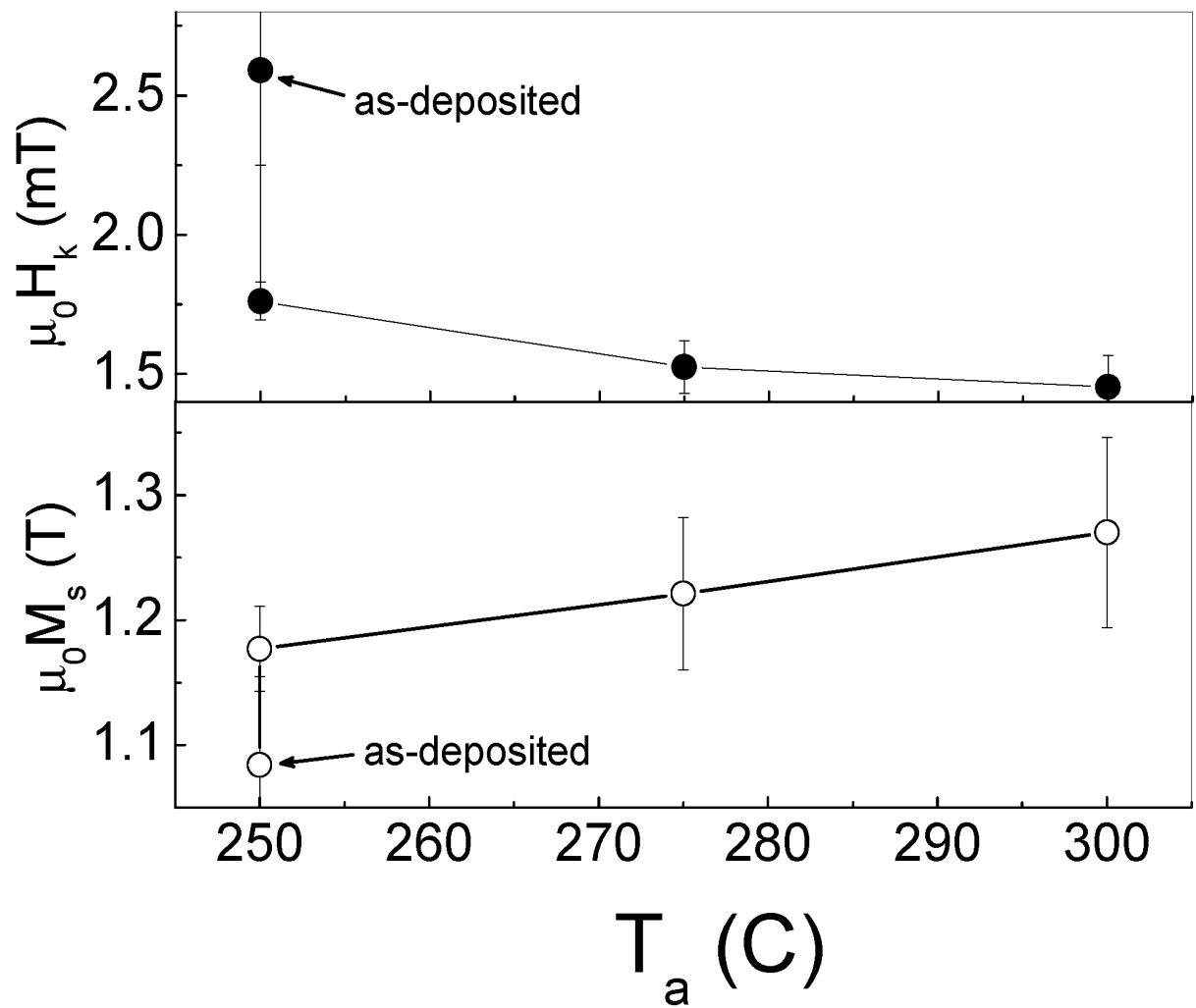

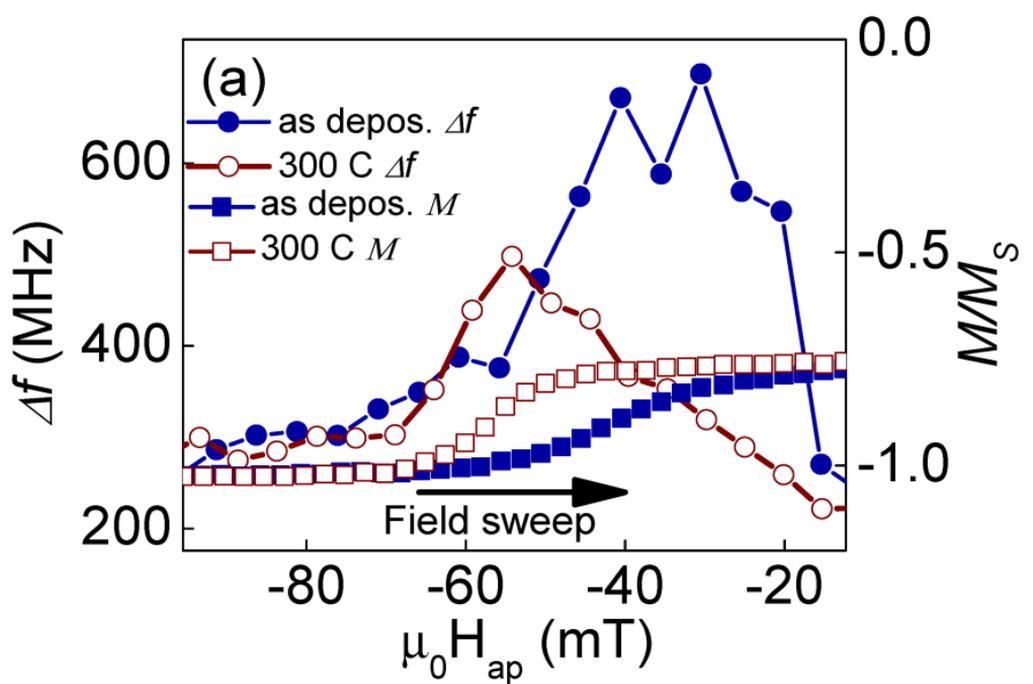

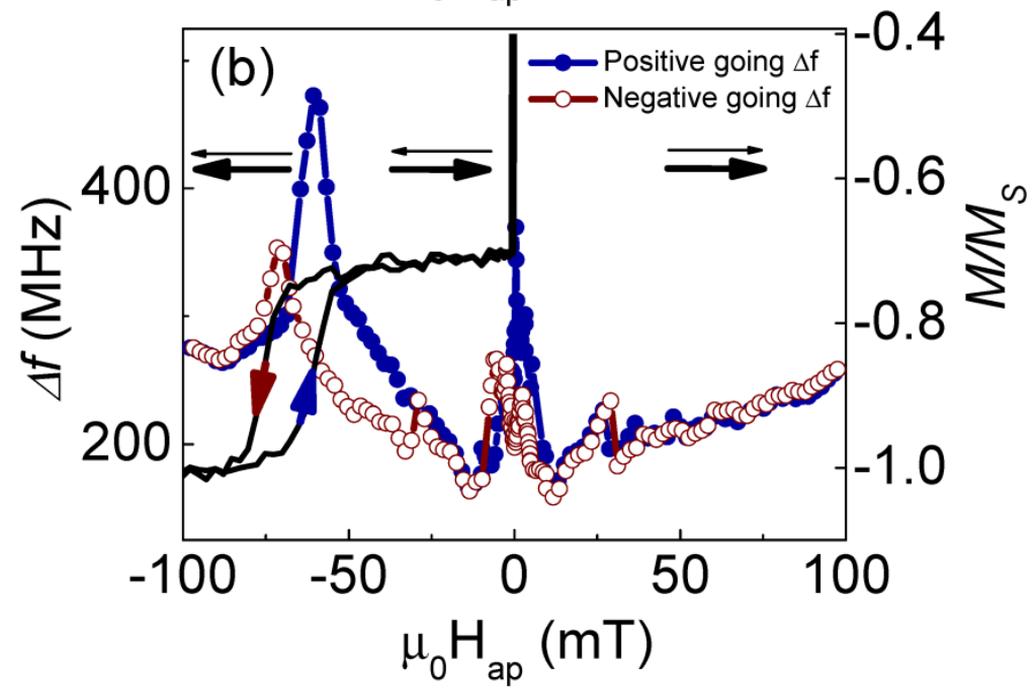